# HRSON: Home-based Routing for Smartphones in Opportunistic Networks

Hooman Abasi
*Faculty of Network Science and Technologies*
University of Tehran
Tehran, Iran
h.abasi@ut.ac.ir

Mostafa Salehi
*Faculty of Network Science and Technologies*
University of Tehran
Tehran, Iran
mostafa_salehi@ut.ac.ir

Vahid Ranjbar
*Faculty of Network Science and Technologies*
University of Tehran
Tehran, Iran
vranjbar@ut.ac.ir

*Abstract*—Opportunistic networks are delay-tolerant networks which enable network connectivity while there is limited access to network infrastructure, such as natural disaster happenings. Since opportunistic networks use store-carry-forward mechanism to deliver data, routing algorithms have significant impact on successful data delivery. Due to the Operating System restrictions, creating an opportunistic network using ad-hoc model is not feasible on smartphones and to the best of our knowledge, none of common zero-knowledge routing algorithms have been proposed for a non-ad hoc communication model. In this paper, we propose HRSON, a zero-knowledge routing algorithm in a self-organizing approach that an opportunistic infrastructure-based WiFi network is built to use smartphones. We have deployed our approach in simulated scenario of working days of people, whom are using smartphones, on Helsinki map comparing to common zero-knowledge routing algorithms. The results show that our solution increases the average delivery-rate and lowers delay and commutation overhead.

## I. Introduction

In the last years, there has been a significant increase in the number of smart devices with high processing power and wireless communication capabilities, such as smart-phones. Accordingly, implementation of opportunistic networks in societies has been more feasible. Opportunistic networks are wireless, delay tolerant networks that enable wireless communication among devices by using mobility as a benefit of carrying messages, where there is no access or limited access to network infrastructure. In opportunistic networks, nodes use store-carry-forward mechanism to deliver messages to destinations instead of end-to-end routing [1], which means a node that receives a message stores it, then carries the message until it would find a suitable node for relaying toward the destination and forwards the message to that node. Since there are no end-to- end paths between nodes due to the mobility of nodes, routing is an important issue in opportunistic networks. Opportunistic networks are used in many areas such as space communication, wild-life monitoring, social applications, cellular traffic offloading, vehicular networks and environments with wireless communications challenges [2]. The advantages of opportunistic networks are two-folded [3]: 1) their infrastructure is more flexible than the cellular networks, and the workload of the network can be alleviated by opportunistic transmissions; 2) the corresponding transmissions are highly energy efficient and low cost.

In wireless networks, nodes are connected either in infrastructure-based mode or ad-hoc mode. In infrastructure-based mode, nodes connect to central nodes with higher capabilities which act as intermediate nodes for communication. In ad-hoc mode, nodes connect to each-other directly and without any intermediate nodes. Nodes in opportunistic networks are usually connected in ad-hoc mode, but due to Operating System restrictions, most of smart-phones do not support ad-hoc mode in wireless communications [4]. In recent years in order to exploit the use of opportunistic networks in smart-phones, self-organizing approaches have been proposed such as WLAN-Opp [4] and SASO [5]. These approaches are based on infrastructure-based mode of WiFi operations in which devices are divided into roles of access points and clients. Access points mainly receive connections and clients can only exchange data through access points. The difference between this mode and common infrastructure-based mode is that any node can be a central nodes and act as an access point.

The majority of the researches on opportunistic networks are related to data dissemination, with focus on routing and mobility models leading to data sharing [6]. Based on the knowledge of the nodes from the network, routing algorithms in opportunistic networks can be divided in two categories of knowledge based and zero knowledge algorithms [3]. In zero knowledge routing algorithms such as Epidemic [7] and SnW [8], nodes have no information of the network such as network topology and mean time between successive meetings of two nodes; while in knowledge-based algorithms such as PRoPHET [9], MAXPRoP [10], SGBR [11] and BubbleRap [12] nodes use this knowledge for routing.

To the best of our knowledge, all of the routing algorithms mentioned above are tested, deployed or simulated in ad hoc mode of opportunistic networks. We propose Home-based Routing of Smartphones in Opportunistic Networks (HRSON), a zero-knowledge routing algorithm to be used in infrastructure-based communication of smartphones such as WLAN-Opp and SASO. This solution is based on an infrastructure-based model of communication among smartphones nodes in which each node takes the role of access point or client and these roles change in accordance with the context of the communication environment. In HRSON, each node can only become an access point if it is in its Home. By this approach we limit data forwarding only in Homes of the nodes. Consider Fig. 1. If a node becomes access point in a non-Home location, since it will leave the area soon, the connections to that node would be terminated and the network needs to be reconstructed. This reconstruction includes scanning for nearby alternative access points and becoming an access point if none is available or connecting to available access points. This process takes considerable time if it would happen for large scale of nodes. By not having this situation as much as possible, this considerable amount of time would be used to transfer data.

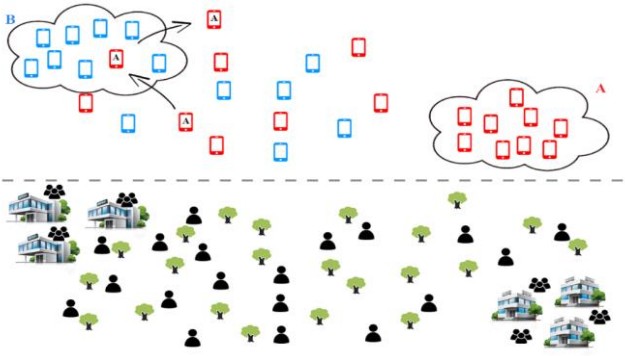

Fig. 1 If a node becomes access point in a non-Home location, and other nodes connect to it, since it will leave the area soon, the created network will disrupt and a time should pass to reinitiate a network

Our solution is simulated in a real-life like condition, based on [13] which simulates working days of human nodes and their movement to their home, office and evening spot, which we refer to as Home in this article. We believe that our approach reduces the overhead time spent for changing roles between access point and client, thus enabling the network to be maintained longer. For more efficiency, we restrict the messages each node can forward, based on binary SnW [8] mechanism, and for evaluation we use the ONE simulator [14]. We compare our approach with known zero knowledge routing algorithms such as Epidemic routing and SnW which we have deployed on SASO [15], the state machine for smart phones' opportunistic connectivity. We evaluate our solution in different situations by changing the simulation parameters such as traffic generation rate, TTL of the packets, number of copies of the messages and number of Homes. The simulation results show that in the most of the considered scenarios our solution is more efficient than mentioned solution in terms of delivery probability, delivery delay and communication overhead ratio, especially when the number of the Homes increase. Overall, we make the following contributions:

- We introduce HRSON, a zero-knowledge routing algorithm for efficient use of opportunistic networks on smart-phones.
- We simulate our solution in a real-life like scenario of people using smart-phones on Helsinki map.
- We evaluate our approach by comparing simulation results of known zero knowledge opportunistic networks routing algorithms in delivery rate, delivery delay and communication overhead ratio.

This paper is organized as follows. Section II discusses related approaches in this literature. In Section III we formalized our system, the problem we want to solve and analyze the solution we propose. In Section IV we evaluate our solution in different scenarios and compare it to well-known existing approaches. Section V concludes the paper.

## II. RELATED WORKS

A significant effort has been made to enhance forwarding protocols in intermittently connected networks in the last years. The objective of message forwarding is to answer the following question: *To whom should a node forward a certain message, in order to maximize the message delivery while minimizing the delay and the number of replicas* [2]. This problem leads to routing algorithms that differ in knowledge that nodes have about the network state such as network topology, node encounters and encounter time between nodes. Based on this, we can divide routing algorithms in opportunistic networks into zero knowledge, and knowledge-based algorithms in which nodes use the information available in the network to forward messages. Some of knowledge-based routing algorithms are followings. PRoPHET [9] that uses delivery probability metrics to calculate how likely a node will be to deliver a message to the destination. MAXPRoP [10] is an extension of PRoPHET with buffer management functionality. In MAXPRoP messages are sorted based on number of hops from their source to the current node and Messages with less hop counts will be sent first. SGBR [11] is a routing algorithm that utilizes social relations of the nodes in data forwarding. In SGBR each node forwards messages to the nodes that are not in its social community unless the receiving node is the destination node. BubbleRap [12] uses local and global centrality to forward messages. In BubbleRap, messages are first sent to the communities with higher global centrality. When the message reaches the destination node's community, it will be forwarded to the nodes with higher centrality until the message reached the destination node.

Zero knowledge algorithms, unlike knowledge-based algorithms, do not use information available in the network. This type of algorithm is applicable when network behavior is unpredictable. It can make few assumptions about future position of the nodes. Zero knowledge algorithms are faster than knowledge-based algorithms and more suitable for networks with large number of nodes [2]. The disadvantage of knowledge-based algorithms compared to zero knowledge algorithms are 1) using knowledge in data forwarding increases communication overhead; 2) learning the knowledge takes time from minutes to several weeks. Epidemic routing and SnW are well known zero knowledge algorithms that are commonly used as a benchmark for evaluation of this type of algorithm. In Epidemic routing, each node forwards receiving messages to all of its neighbors, which will also repeat this process until either the message is delivered to the destination node, or Time to Live (TTL) of the message is passed. SnW (Spray and Wait) limits the number of messages each node can forward to other nodes. SnW consist of two phases: Spray phase in which L copies of the message are spread to distinct nodes, and waiting phase in which receiving nodes of the copies carry them until they find the destination node. Table 1 summarizes representative routing algorithms in opportunistic networks.

All the algorithms mentioned above are using ad-hoc mode of Opportunistic Networks. Except for Epidemic routing, which has been used in WLAN-Opp and SASO, none have been experimented or proposed to be used in opportunistic networks of smartphones deployed on infrastructure-based connectivity. Since ad-hoc mode is not supported in smart-phones, recently WLAP-Opp and SASO proposed state-machines which exploit the infrastructure mode to the benefit of implementing Opportunistic Networks for smart-phones. In WLAN-Opp [4], based on contextual information, each node may become an access point and other nodes become its clients. In order to extend the network, so all of the nodes would be able to communicate, a state-machine is proposed in which nodes change their roles from access point to client or vice versa. WLAN-Opp simulation results show that their approach is more efficient than ad-hoc mode in terms of delivery probability, delay and throughput. SASO improved WLAN-Opp by defining multiple WiFi channels used for overlapping access points in order to increase data transfer speed. Both WLAN-Opp and SASO use epidemic data forwarding. In these approaches, nodes can become an access point on every location on the simulation map. We assume that each node visits some locations more frequently, or stays in longer. We refer to these locations as Homes of node. Consider a situation in which a node becomes an access point while it is not in its Home becomes an access point and other nodes become its clients. Since the node is not in its Home location, soon it will leave the area, disrupting the existing network. This situation is demonstrated in Fig. 1. After the access point node leaves the network it takes time to rebuild it and through that time no transmission occurs, lowering the chance of successful delivery.

TABLE 1 SUMMARY OF THE REPRESENTATIVE ROUTING ALGORITHMS IN THIS ARTICLE

| Algorithms | | Tested in wireless connectivity environment | | Optimized for infrastructure-based connection for smart-phones |
|---|---|---|---|---|
| | | Ad hoc | Smartphone's infrastructure based | |
| Zero knowledge | Epidemic | ✓ | ✓ | ✗ |
| | SnW | ✓ | ✗ | ✗ |
| Knowledge based | PRoPHET | ✓ | ✗ | ✗ |
| | MAXPRoP | ✓ | ✗ | ✗ |
| | SGBR | ✓ | ✗ | ✗ |
| | BubbleRap | ✓ | ✗ | ✗ |

### III. HRSON

We propose a zero-knowledge routing algorithm for opportunistic networks built with smartphones. Our approach is based on an infrastructure-based model of communication between nodes, in which each node takes the role of access point or client in accordance with the context of the communication environment. This model is built to enable opportunistic network communications for smartphones. In order to implement HRSON, we considered a five-layered architecture shown in Fig. 2, namely simulation map and routes, movement model, traffic generation, opportunistic network state-machine and routing algorithm. We describe these layers in following.

#### A. Simulation Map and Routes

The first layer consists of the simulations map and the routes on which nodes move. We have used Helsinki's downtown map which is obtained from OpenJump1 software and can be seen in Fig. 3. The reason of using this layer is to implement our approach in an environment similar to the real world, as much as possible. In our simulation, routs, buildings, streets, bus stops and evening spots match the real world.

#### B. Movement Model

The second layer is for the movement model of the nodes. Movement models determine the way nodes move on the map. For the simplicity, many approaches use random movement models. In order to make the simulation similar to real world conditions as much as possible we used Working Day Movement Model (WDMM) [13]. WDMM is a synthetic movement model that simulates working days of people, similar to the real-world movement traces. In WDMM, each node has three locations including its house, office and evening spot. Every day at 8:00 AM, nodes move from their houses to their offices by buses or by cars and work for 8 hours. Nodes also have movements between different segments of their office. At the end of the working day, nodes either move back to their houses or to an evening spot. Either way, after they reach their houses, they stay there until 8:00 AM the next day.

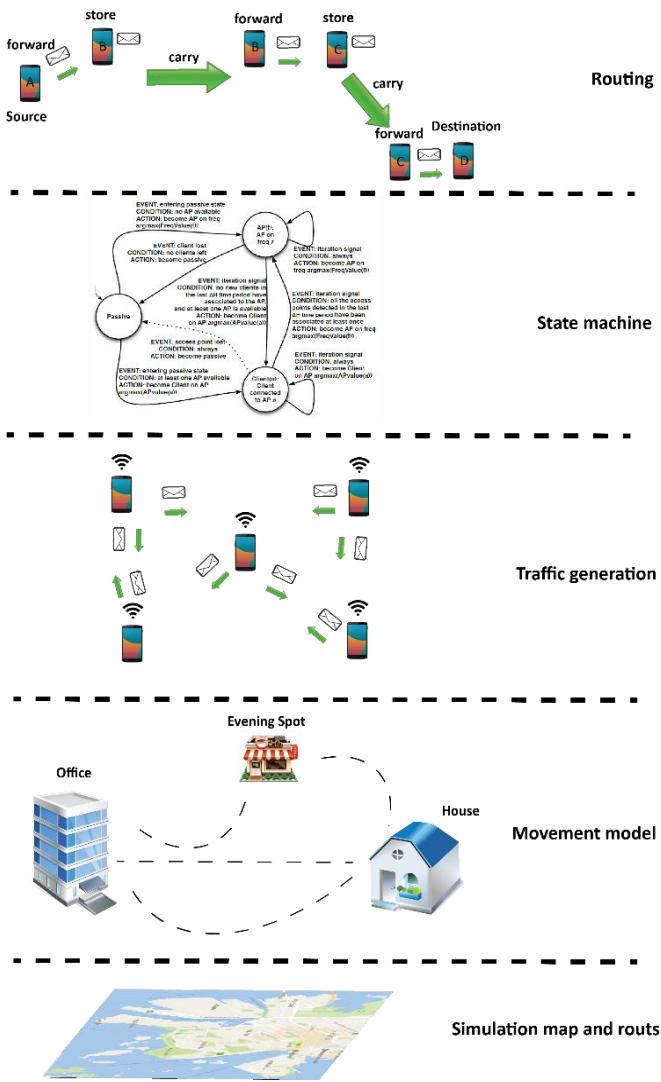

Fig. 2 Layers considered in implementation of HRSON (Details of the state machine can be found at [5])

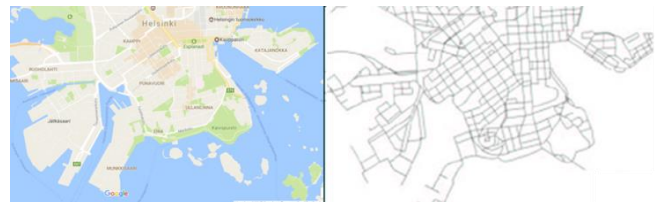

Fig. 3 Helsinki on Google map and simulation map

---

1 http://www.openjump.com

*C. Traffic Generation*

The third layer is the traffic generation of the nodes. In order to evaluate HRSON in different loads of traffic and free increase or decrease of traffic generation rate, we have used random traffic generation model. Each generated packet has a random source and destination, generation time and size.

*D. State Machine*

The forth layer is the state machine which enables opportunistic network for smartphones. We have used SASO [5] since it is the best choice available among the existing state machines. In the state machine there are three roles for each node: 1) Access point which is a central node that other nodes connect to it. Access points are intermediate nodes between their clients. 2) Client that is the role of a node which connects to available access points; and 3) Idle in which nodes decide to either become an access point themselves or connect to available access points. The state machine enables a state transition between these three roles, so that all of the nodes would have the opportunity to visiting each other. The details of the state machine can be seen in [5].

In order to implement our approach, we consider the timing parameters in [5] for real mobile devices operations as following: 1) the time for scanning nearby devices: $t\_scan = 5\ seconds$ 2) the resting time between scans: $t\_rest = 1\ second$, 3) the time to become access point from any state: $t\_ap = 1\ second$ 4) the time to become client of an access point: $t\_connect = 5\ seconds$. Using these parameters, the time to initiate a network is calculated as follows:

$$t\_net\_initiate = t\_scan + t\_ap + t\_scan + t\_connect = 16\ seconds$$

$t\_net\_initiate$ shows the steps to build a network, when there is no access point available. These steps are followings:

- Scan for nearby devices in order to locate an access point.
- Since no access point is available, a node becomes access point based on parameters in [15].
- Another $t\_scan$ is spent by other nodes than the access point in order to locate it.
- $t\_connect$ is spent for other nodes to connect to the new available access point.

If a situation like the one demonstrated in Fig. 1 occurs, the network will fall apart soon. In order to rebuild the network, another process of network initialization should begin. If there are access points available, this process will take $t\_net\_reinitiate1$ amount of time:

$$t\_net\_reinitiate1 = t\_scan + t\_connect = 10\ seconds$$

$t\_scan$ is spent for nodes to find available access points after the network falls apart, and after that $t\_connect$ to connect to the available access point. On the other hand, if there are no access points available after network falls apart, the amount of time to rebuild the network will be calculated as:

$$t\_net\_reinitiate2 = t\_net\_initiate = 16\ seconds$$

$t\_net\_reinitiate1$ shows that, if the present access point leaves the network and there are other access points available in adjacency, the time to rebuild a network equals 10 seconds and $t\_net\_reinitiate2$ shows that, if the current access point leaves the network and there are no other access points available in adjacency, the time to rebuild a network equals 16 seconds. These time values are actually the time that there is no data forwarding among parted nodes in the disrupted network. If we consider the wireless transmission speed to be 5 MBps, for the first scenario the network transmits 50 MBs less data and for the seconds scenario this amount equals 80 MBs. With larger scale of nodes this amount would also increase significantly.

*E. HRSON Routing*

We propose a zero knowledge Home-based routing algorithm that can be used in infrastructure-based opportunistic network connections of smart phones. In our approach, each node can only become an access point if it is in one of its Homes. We also use the binary SnW mechanism to limit the number of messages in the network. We believe that our approach can improve the average packet delivery rate, lower the average latency and lower the communication overhead compared to other well-known zero knowledge routing algorithms. The pseudo code for the proposed algorithms is presented Procedure 1.

The steps of HRSON is as followings:

- The node receives or generates a message. The message has copying limitation $L$.
- While the message is not delivered and the node is connected to a network, if $L \geq 2$, it sends L copies to $L$ encountering nodes. After that, the number of copies allowed to forward for receiving nodes becomes $L / 2$.
- If $L < 2$, the node waits until it reaches the destination node. Otherwise, the forwarding routine continues.
- If the node is not connected to a network:
  - If there are access points available nearby, it connects to the fastest access point.
  - If there are no access points available, if the node is in its Home, it becomes an access point itself.

In section four, we will discuss simulation setup and evaluation of HRSON.

IV. SIMULATION SETUP AND EVALUATION

To test our proposed approach and compare its performance with other approaches we used the ONE simulator [14]. The results of each experiment are averaged among 32 Mont Carlo simulations of the same experiment with different random number seeds for 5 simulated working days. The Evaluation metrics are delivery probability, average latency and communication overhead. The parameters considered for the simulation setup is based on the five-layer architecture depicted in Fig. 2. In following we describe these parameters.

## A. Parameters of the Simulation Map and Routes

The simulation map and routes are extracted from Google map using OpenJump software. In Fig. 3 you can see that every section of the real map matches the simulation map. The size of this map is approximately $1750 \times 2125$ m$^2$.

## B. Parameters of the Movement Model

In order to deploy our approach in a scenario similar to the real world, we have used WDMM. An instance of a nodes' movement in WDMM can be seen on the movement model layer of Fig. 2. The process of WDMM was explained in section 4, part B. The parameters of the movement model used in simulation setup can be seen in Table. 3.

Prior to the simulation initialization, nodes are randomly place on the map, divided into five groups of A, B, C, D and E. These groups are not social groups, but they are group of nodes in the same locations of the map. House, office and evening spot of every node is in the location of its. The reason of this segmentation is that for most people these locations are close to each other. For instance, it is common that people choose their houses near their work for less more comfortable commute. We have considered three main segments called A, B and C, and two overlapping segments called D and E, respectively representing the overlap of A and B, and A and C. Number of nodes in groups in A, B, C, D and E are respectively 325, 275, 300, 50 and 50.

PROCEDURE 1. HOME-BASED PROCEDURE FORE BOCIMING AN ACCESS-POINT

```
1. // Number of copies allowed for each message
2. L = n
3. /* For each message, do the forwarding until the receiver is the
      destination node. */
4. for each packet
5.    while isDeliverd () == false then
6.       // Until the node is connected to a network, do the forwarding.
7.       while nodeIsConnected == true then
8.          If encounterDestinationNode == true then
9.             finalTransfer ()
10.            delivered = true
11.            stopTransferring ()
12.         // message copy limitation mechanism
13.         else if copy >= 2 then
14.            sendToUnseenNeighbors ()
15.            L = L/2
16.         end if
17.         // search for a faster network
18.         scanForFasterAp ()
19.         if betterApAvailable () == true then
20.            connected = false
21.            break
22.         end if
23.      end while
24.
25.      while nodeIsConnected == false then
26.         scanForAp ()
27.         if apAvailale == true then
28.            becomeCliente ()
29.            connected = true
30.            Break
31.         else if isAtHome == true then
32.            becomeAp ()
33.            connected == true
34.            Break
35.         end if
36.      end while
37.   end while
38. end for
```

## C. Parameters of Traffic Generation

We have used random traffic generation in the simulation setup. Each message has a random source and destination, size, generation time and TTL. Traffic generation is a variable parameter. Message size is constant and is between 500 Kbytes and 1500 Kbytes.

## D. Parameters of the State Machine

In order to implement opportunistic network on smartphones we have used SASO. The parameters of the state machine are based on Google Nexus 10 with Android Operating System which has been used in this research. These parameters can be seen on Table 2.

TABLE. 3 PARAMETERS OF THE MOVEMENT MODEL

| Input Parameter | Amount | Description |
|---|---|---|
| Number of houses | Variable | Overall number of houses on the map |
| Number of offices | Variable | Overall number of offices on the map |
| Number of evening spots | Variable | Overall number of evening spots on the map |
| Working hours | 28800 seconds | The amount of time nodes spend in office |
| Office wait time | 10 - 100000 seconds | The amount of time each node pauses after a movement in the office |
| Office size | 10000 m$^2$ | Size of the offices in |
| Evening activity probability | 0.5 | The probability of going to an evening spot after work |
| Size of evening spot's groups | 1 - 3 | Number of people in a group that gather in an evening spot (friends) |
| Evening spot wait time | 3600 - 7200 seconds | The amount of time nodes stay in evening spots |
| Own car probability | 0.5 | The probability of node moving between places with a car |
| Bus wait time | 10 - 30 seconds | The amount of time a bus waits in a bus stop |
| Car and bus movement speed | 7 - 10 m/s | The speed of the node if it would use buses or cars for moving between places |
| Node movement speed | 0.8 – 1.4 m/s | The speed of the node while moving on foot |

TABLE 2. PARAMETERS OF THE STATE MACHINE

| Input parameter | Amount | Description |
|---|---|---|
| Buffer size | 100 Mbytes per second | The storage each node has for carrying the messages as a relay node |
| Wireless transmission speed | 5 Mbytes per second | Transmission speed based on IEEE 802.11 standards |
| Communication range | 20 m | The range of wireless communication |
| Scan wait time | 1 second | The amount of time between scans for nearby access points. |
| Scan time | 5 seconds | The scanning time to search for nearby access points |
| Become AP time | 1 second | The time that takes the node to become an access point from any state |
| Become client time | 5 seconds | The time that takes the node to become client from any state |
| Number of none-overlapping network | 5 | Number of ranges of frequency that do not overlap for nearby networks resulting in higher transmission speed |

## E. Evaluation Scenarios

In order to evaluate HRSON on different conditions, we have considered four scenarios. We compare the results of these scenarios with Epidemic and SnW routing algorithms in delivery probability, average latency and communication overhead. Table 4 shows the variable parameters used in each scenario. As you can see, in each scenario three parameters are constant and one is variable. The purpose to vary each parameter is followings:

- Traffic generation: Testing the behavior of HRSON in low and high rates of traffic in the network compared to other algorithms.
- TTL: If TTL is infinite, eventually all the messages will reach the destination. But in real world each message has a life time. An efficient routing algorithm should deliver higher rates of messages with less latency.
- Number of copies: Since this parameter is one of main features of HRSON, we vary it to see the behavior of our approach based on it.
- Number of Homes: HRSON is a Home-based routing algorithm. We vary number of Homes to see this impact on HRSON.

The evaluation metrics are delivery rate which shows the ratio of delivered to generated packages, average delay which is the time between message creation and message delivery and communication overhead which includes the effort between nodes, such as control and signaling data, to achieve a reliable transmission. Since communication overhead of SnW and HRSON are almost 33 times smaller than Epidemic in all scenarios, we only compare HRSON with SnW in this case. The reason of this distance is that HRSON and SnW limit the number of message copies and there are much less message forwarding in these two.

***Varying traffic generation.*** In this scenario we have varied traffic generation to create one message every 75 to 100 seconds, 50 to 75 seconds, 25 to 50 seconds and 10 to 25 seconds. These traffic models respectively generate 2661, 3742, 6285 and 13646 messages in the network. If you take a look at Fig. 4, you can see that with the increase of traffic generation in the network, HRSON and SnW almost show the same results in all of evaluation metrics. Although, HRSON is slightly better. Both have a slight increase in delivery rate, then become almost constant and both have a decrease in average latency. Delivery rate of Epidemic decreases with a high slope because of network saturation and hence decrease in communication speed and loss of more messages. Average latency of Epidemic is much higher than other two algorithms. The communication overhead increases in both HRSON and SnW because of the increase in message generation and hence more communications among nodes.

***Varying message TTL.*** In this scenario, we have varied message TTL with numbers 6, 12, 18 and 24. Fig. 5 shows that with the increase of message TTL, delivery rate of all three algorithms increase and after 18 TTL it decreases for Epidemic due to network saturation and hence decrease in communication speed and loss of more messages. We can see that delivery rate of HRSON and SnW are close to each other, but their distance increases slightly with the increase of message TTL. With increase of message TTL, the average latency increases for all algorithms; Although, HRSON and SnW are very similar and have less latency than Epidemic. The communication overhead of HRSON and SnW decrease with the increase of message TTL. This is because of that less processing power is spent for dropping messages from the nodes' buffer. HRSON has slightly less overhead than SnW.

***Varying the number of message copy limits.*** In this scenario, we have varied the allowed number of copies for each message to be spread in the network with numbers 4, 8, 12, 16 and 20. Message copy limitation is a feature of HRSON, so our approach should have better results compared to other ones. On Fig. 6, you can see that with the increase of message copy limitation delivery rate of HRSON and SnW increases and for Epidemic, since copy change doesn't have an effect on, it is constant.

You can see that delivery rate of HRSON is always higher than SnW and their distance increase with the increase of message copy limitation. The average latency for HRSON and SnW decrease and for Epidemic it is again constant. Although HRSON has less overhead, the communication overhead of HRSON and SnW increases since there will be more communications when there are higher message copy limits. We can see that with increase of copy limitation, message delivery increases but there is also an increase in communication overhead. We recommend the amount of 12 for better message delivery and also less communication overhead.

***Varying number of Homes.*** In this scenario we have varied number of Homes for the nodes. Since HRSON is Home-based and data forwarding is restricted at Homes, this is the main feature of our approach. In this scenario, number of houses is 203. This number is constant and the locations are based on dataset of WDMM. To evaluate HRSON with different number of Homes, we have varied the number of offices and evening spots and used random specification of these Homes on the map.

Table 5 shows the number of Homes for each time that this scenario has been deployed. If you take a look at Fig. 7, you can see that with increase in the number of Homes, all of the algorithms have a decrease on delivery rate due to less node density and hence less communication opportunity; Although HRSON exceeds SnW significantly, and with a slight distance is better than Epidemic. The average latency of HRSON is lower than both SnW and Epidemic, and has high distance with Epidemic. The communication overhead of HRSON is less than SnW and is almost constant. The recommended amount for number of Homes for more efficiency of HRSON is 2 Homes per number of nodes, which is 500 Homes for 1000 nodes in this simulation scenario.

TABLE 4. VARIABLE PARAMETERS OF THE SIMULATION

| Scenario | Traffic generation | Message TTL | Number of copies | Number of Homes |
|---|---|---|---|---|
| First | One message every 10 to 50 seconds | 24 hours | 10 | Variable |
| Second | One message every 10 to 50 seconds | 24 hours | Variable | 257 |
| Third | One message every 10 to 50 seconds | Variable | 10 | 257 |
| Forth | Variable | 24 hours | 10 | 257 |

TABLE 5. VARYING NUMBER OF HOMES

| Experiment number | Number of houses | Number of offices | Number of evening spots |
|---|---|---|---|
| One | 203 | 50 | 10 |
| Two | 203 | 150 | 30 |
| Three | 203 | 250 | 50 |
| Four | 203 | 350 | 70 |
| Five | 203 | 450 | 90 |

The cost for the efficiency of HRSON is that it stores messages in its buffer almost 25 times longer than Epidemic, due to the message copy limitation feature. This also happens for SnW. Although, this amount of time is slightly less for HRSON.

## V. CONCLUSION

In this paper we introduced a zero knowledge routing algorithm for opportunistic networks using smart phones, namely HRSON, relying on a Home-based mechanism of data forwarding with limitation on number of message forwarding. HRSON is a routing approach in which each node can operate in an infrastructure-based state machine of wireless connection between smart phones and forward data only in locations with high frequency of visits, namely Homes of the nodes. We implemented HRSON on a four-layer architecture of simulation map, movement model, traffic generation and state machine of smartphones. We evaluated the performance of the proposed approach in a real-life like simulation of people's working days within a week in large scale of number of nodes and a non-random movement model within a map from the real world in different conditions. We compared our approach with well-known zero knowledge routing algorithms of opportunistic networks. The simulation results show that this work surpasses these algorithms in delivery rate, average latency and communication overhead in most of the cases and this superiority is more significant with the increase in number of Homes and message copy limit, which both are main features of the algorithm.

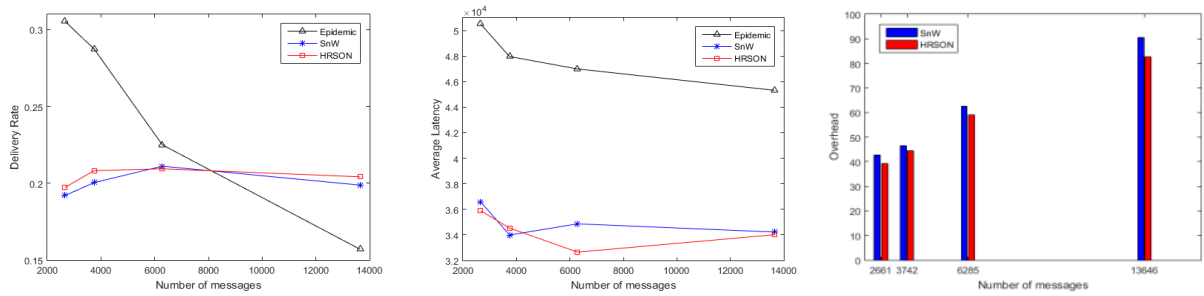

Fig. 4 Varying the number of messages. The plots show delivery rate, average latency and communication overhead.

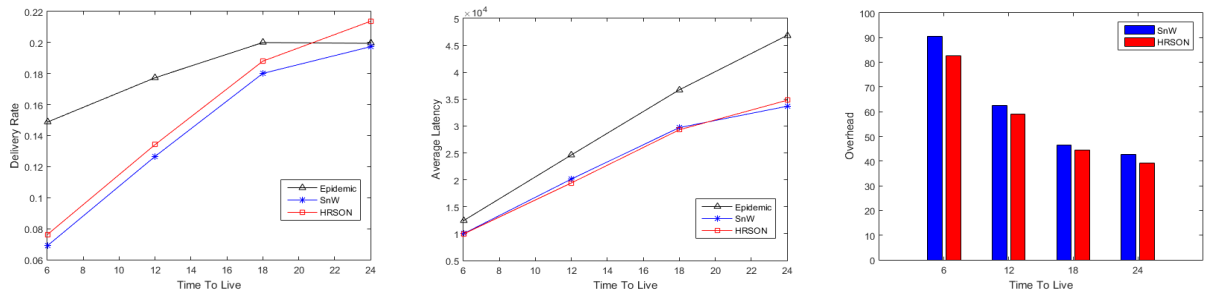

Fig. 5 Varying TTL of the messages. The plots show delivery rate, average latency and communication overhead.

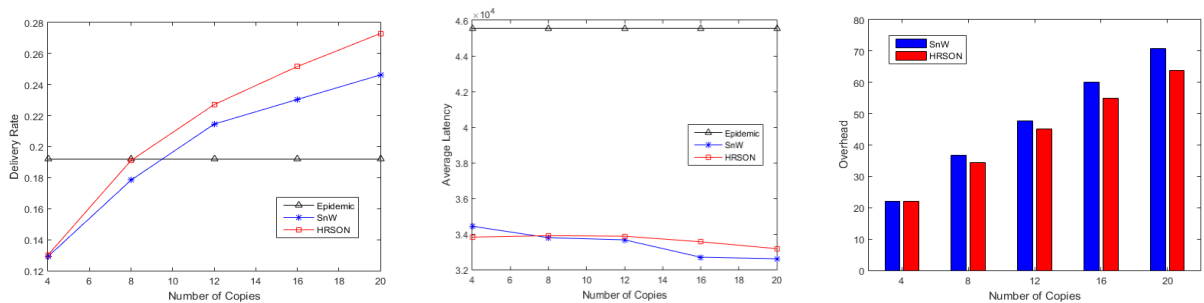

Fig. 6 Varying the number of copies for each message. The plots show delivery rate, average latency and communication overhead.

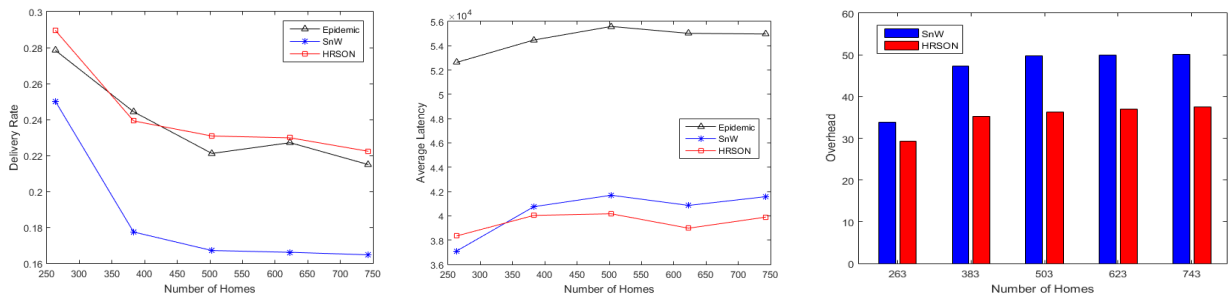

Fig. 7 Varying the number of Homes. The plots show delivery rate, average latency and communication overhead.

In the future, we expect that it will be possible to implement HRSON on real smartphones. For this purpose, we should consider existing middle-wares such as ShAir [15] and Haggle [16]. In the latter, the devices can connect through available infrastructures, cellular network or opportunistic network, according to their conditions and their need. To the best of our knowledge none of mentioned middle-wares have implemented a routing module. We hope be able to create a routing module for higher package delivery with lower delay for real smart phones.